\documentclass[a4paper]{jpconf}
\usepackage{graphicx}
\usepackage{hyperref}
\input{babarsym}
\usepackage{xcolor}
\usepackage{lineno}
\selectcolormodel{rgb}
\definecolor{LHCb dark}{rgb}{0.0000,0.3412,0.6549}%From LHCb logo
\definecolor{UC red}{rgb}{0.8196,0.1176,0.2314} % from logo
\definecolor{brickred}{rgb}{0.8, 0.25, 0.33}

\def\swave {{\ensuremath{\cal S}}-wave\xspace}
\def\pwave {{\ensuremath{\cal P}}-wave\xspace}
\def\dwave {{\ensuremath{\cal D}}-wave\xspace}

\begin{document}
\title{Model-independent partial wave analysis using a massively-parallel fitting framework}

\author{L Sun$^1$, R Aoude$^2$, A C dos Reis$^2$, M Sokoloff$^3$}
\address{$^1$ School of Physics and Technology, Wuhan University, Wuhan, Hubei Province, 430072 China }
\address{$^2$ Centro Brasileiro de Pesquisas F\'{i}sicas (CBPF), Rio de Janeiro, 22290-180 Brazil }
\address{$^3$ University of Cincinnati, Physics Department, ML0011, Cincinnati OH 45221-0011, USA}

\ead{sunl@whu.edu.cn}

\begin{abstract}
The functionality of  GooFit, 
a GPU-friendly framework for doing maximum-likelihood fits, 
has been extended 
to extract model-independent \swave amplitudes in
three-body decays such as $D^+ \to h^+h^+h^{-}$. 
A full amplitude analysis is done where the magnitudes and 
phases of the \swave amplitudes %(or alternatively, 
%the real and imaginary components), 
are anchored at a finite number of $m^2(h^+h^{-})$ control points, 
and a cubic spline is used to interpolate between these points. 
The amplitudes for \pwave and \dwave intermediate states are modeled as 
spin-dependent Breit-Wigner resonances. 
GooFit uses the Thrust library, %to launch all kernels,
    with a CUDA backend 
for NVIDIA GPUs and an OpenMP backend for threads with conventional 
CPUs. 
Performance on a variety of platforms is compared. 
Executing on systems with GPUs is typically a few hundred times faster than 
executing the same algorithm on a single CPU.
\end{abstract}

\section{Introduction}
%Dalitz plot analyses~\cite{dalitz} provide a method  to study the
%quantum mechanical amplitudes underlying the
%dynamics of three-body decays.
%Such decays of charm mesons are
%expected to proceed predominantly through intermediate
%quasi-two-body modes.
%This is the experimentally  
%observed pattern. 
%Amplitude analyses determine the relative contributions
%of processes that contribute to the observed
%three-body final states. 
%In addition, since the intermediate
%quasi-two-body modes are dominated by light
%quark meson resonances, new information on light meson
%spectroscopy can be obtained. 
%In particular, the Model Independent Partial Wave
%Analysis (MIPWA) technique~\cite{mipwa}, described below,
%promises to improve our understanding
%of the structure of \swave resonances, a long-standing 
%puzzle in light meson spectroscopy.

The physics of scalar mesons below 2~\gev is a long-standing 
puzzle in light meson spectroscopy~\cite{pdg2016}, 
this is partly due to the fact that these mesons often 
have large widths and overlap with each other, 
which make them hard to model. 
Contributions from Light scalars can be extracted
by using Dalitz plot analyses~\cite{dalitz} 
in three-body hadronic decays of charm mesons.
%These decays often proceed through intermediate quasi-two-body 
%modes. 
First developed by the E791 Collaboration~\cite{mipwa}, 
the Model Independent Partial Wave
Analysis (MIPWA)
technique  extracts \swave amplitudes
in the Dalitz plot analysis
with no  assumption about the nature of the \swave. 
To pin down its large numbers of free parameters, 
the MIPWA technique requires the large samples
of three-body decay events that have 
become increasingly available at the $B$-factories and the LHC. 
Notably, the LHCb experiment has recorded charm decays 
with sample sizes exceeding any previous experiment by more than
an order of magnitude, 
offering a unique opportunity
to study \swave structures with unprecedented levels of precision. 
Analysing very large statistics samples requires 
disproportionately more computing power. 
Running all the calculations in a single Central Processing unit (CPU)
thread would take a prohibitively long time. 
%Even though modern
%multi-core CPUs are capable of thread-level parallelism, the number of 
%parallel threads is still significantly limited.

Originally, a  Graphics Processing Unit (GPU) was a specialised 
electronic circuit designed to rapidly 
manipulate and alter memory to accelerate the creation of 
images for output to a video display. 
The
highly parallel structure makes GPUs more effective than 
CPUs for algorithms where large blocks 
of data are processed in parallel.  
Today, the functionality of GPUs has been extended to enable
highly parallel computing for scientific and other more
general applications.
An open-source framework called GooFit~\cite{goofit}
has been developed to exploit the processing power of these GPUs
for parallel function evaluation, particularly in the
context of maximum likelihood fits.  
This paper reports an extension of the GooFit framework 
to support MIPWA of a three-body decay
with 
vastly reduced processing time. 

\section{MIPWA method}
Essentially all studies of three-body hadronic $D_{(s)}$ decays 
employ the same technique: the unbinned maximum likelihood
fit of the Dalitz plot, in which the quantum mechanical matrix element
governing the decay
is represented by a coherent sum of %phenomenological
amplitudes~\cite{pdg2016}. 
These amplitudes correspond to the 
possible intermediate states in the decay chain 
$D_{(s)}\to R h_3, R\to h_1 h_2$ ($h=K,\ \pi$). 
The amplitudes are grouped according to the orbital angular momentum $L$ 
of $R$ and $h_3$ in the rest frame of the $D_{(s)}$, 
and the total amplitude is

\begin{equation}
A(s_{12},s_{13}) = \sum_L \sum_k c_k^{L}A_k^{L}(s_{12},s_{13}) \, ,
\end{equation}
where $s_{12(13)}\equiv m^2(h_{1} h_{2(3)})$. 
The amplitudes $A_k^L$ are weighted by constant complex coefficients $c_k^L$, 
and the series is truncated at $L = 2$. 
For a resonance with 
non-zero spin, 
the amplitude $A_k$ is most often 
described using a relativistic Breit-Wigner
function multiplied by a real spin-dependent angular factor~\cite{cleobw}. 
For the \swave, two qualitatively
different approaches exist. 
In the Isobar 
model, the \swave is treated as a sum of a constant non-resonant term and 
Breit-Wigner functions for the scalar resonances. 
While the Isobar model provides reasonably good descriptions for 
narrow resonances, it fails to describe the overlap of broad resonances. 
Additionally, the physical interpretation of the constant non-resonant term
is problematic.

To address these issues, the MIPWA describes the ensemble of scalar 
components using a purely phenomenological set
of parameters derived from the data. 
The $s_{12(13)}$ mass spectrum is divided
into $N-1$ slices with $N$ boundary points separating the slices
and at the two ends of the spectrum. 
The \swave is represented
by a generic complex function $A_0(s) = a_0(s) e^{i\phi_0(s)}$.
At each of the  $k$ boundary points, $A_0(s=s_k) = a_k e^{i\phi_k}$
where  $ a_k $ and $ \phi_k $ are real parameters. 
Between the $ N $ boundary points, the \swave 
is parametrised by a cubic spline~\cite{cubic} in the complex plane.
The set of \{$a_k, \phi_k$\} are free parameters, along
with the coefficients $ c_l^L $ of the higher spin terms, are determined 
in the MIPWA fit. 
The large data sets studied, along with the large number
of parameters ($ 2N $) required to describe the \swave,
make maximising the likelihood a computationally intensive
problem.
Interpolating between the boundary points leads to correlations
between the parameters, and hence to non-linear behavior.
%This makes the problem especially interesting.

\section{MIPWA method with GooFit}
GooFit provides an interface to allow probability density functions (PDFs)
to be evaluated in parallel, using either GPUs or multicore CPUs as back-ends. 
While the original intention of GooFit was to utilise the massive 
computational
power of NVIDIA GPUs based on the proprietary Compute
Unified Device Architecture language (CUDA)~\cite{cuda},
the Thrust parallel algorithms library~\cite{thrust} %developed by NVIDIA
also supports an OpenMP backend for conventional CPUs. 

GooFit has been used to perform a number of amplitude analyses,
for example that of 
Ref.~\cite{pipipi0}.
For a time-integrated Dalitz plot analysis, GooFit provides
the {\tt DalitzPlotPdf} class to model the Dalitz plot PDF. 
The {\tt DalitzPlotPdf}
object contains a list of {\tt ResonancePdf} objects to describe 
resonant amplitudes as well as a constant nonresonant amplitude. 
The original GooFit package  provided only 
the Isobar model to parameterise the \swave. 

The {\tt ResonancePdf} class has been extended to add support for the MIPWA 
method. 
The entire \swave is treated as a single resonance ({\tt ResonancePdf} 
object), with a set of 
free parameters \{$a^k_0, \phi^k_0$\} to be determined in the fit.

\section{MIPWA in $D^+\to K^+K^+K^-$ decay~\footnote{Charge conjugation is implied throughout.}}
The $D^+\to K^+K^+K^-$ decay offers a 
good 
opportunity to study the $K^+K^-$
\swave amplitude directly. 
LHCb collected a large  $D^+\to K^+K^+K^-$ data sample 
from $2.1~\invfb$ of $\sqrt{s} = 7$~\tev $pp$ collisions
recorded by the experiment during 2012. 
About 100K candidates were selected 
with a signal purity of 90\%. The resonant structure of the $K^+ K^-$
\swave amplitude was studied for the first time using an Isobar model,
and the 
results were first presented during the CHARM 2016 workshop~\cite{KKKisobar}.
This analysis indicates that the \swave component accounts for about 90\% 
of the decay rate, while the \pwave ($\phi(1020)$ resonance)
makes up the rest.

\begin{figure}%[h]
%\begin{minipage}{14pc}
\includegraphics[width=\textwidth]{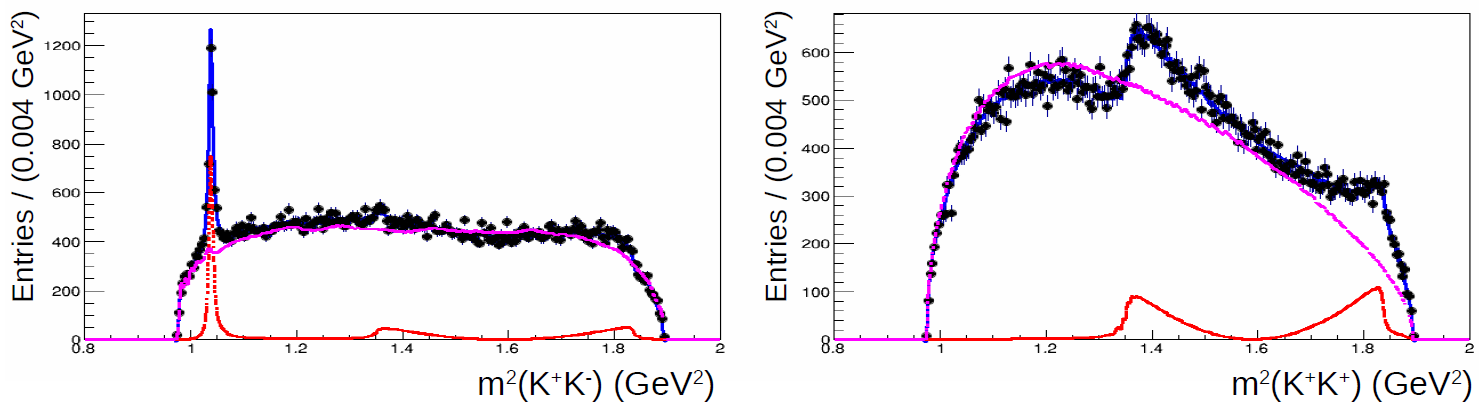}
    \caption{\label{fig:fitproj} The projections of $m^2(K^+ K^-)$ (left) 
    and $m^2(K^+ K^+)$ (right) from $D^+\to K^+K^+K^-$. In each plot, the toy MC signal events (points with error bars ) are shown together with 
    the total fit (blue line), $\phi$ resonance (red line), and \swave (magenta line) 
    determined from the MIPWA.}
\end{figure}

\begin{figure}%[h]
%\begin{minipage}{14pc}
\centering
\includegraphics[width=0.8\textwidth]{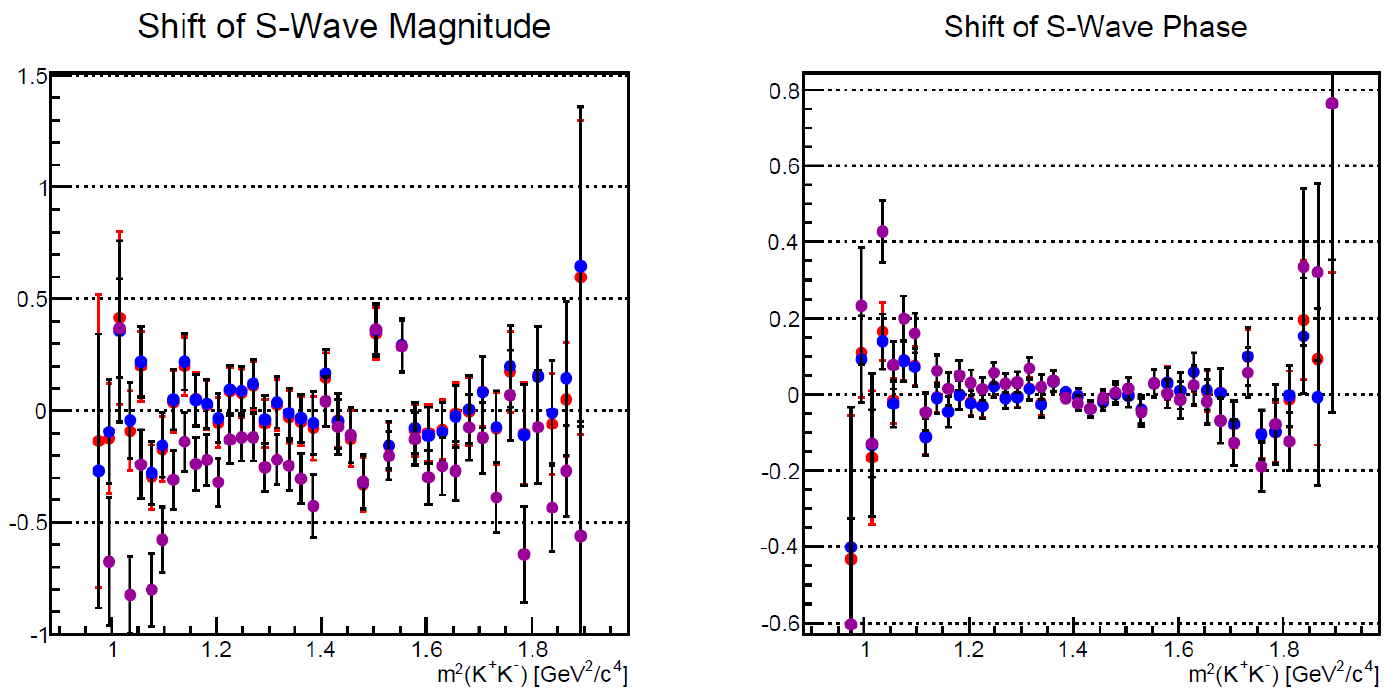}
\caption{\label{fig:swaveshift} Differences between the fitted 
    values and input ones for each boundary point. Shown are set of points with
    different grid spacings for the normalisation: 0.01~${\gev}^2$ (purple),
     0.004~${\gev}^2$ (red), and 0.001~${\gev}^2$ (blue).}
%\end{minipage}\hspace{2pc}%
%\begin{minipage}{14pc}
%\includegraphics[width=14pc]{name.eps}
%\caption{\label{label}Figure caption for second of two sided figures.}
%\end{minipage} 
\end{figure}

A specific goal of the work presented here is to
provide a tool
to extract the $K^+ K^-$ \swave by performing an MIPWA on 
the same LHCb data sample. 
In this case, the Dalitz plot is symmetrised along the 
two axes of $m^2(K^+K^-)$. 
The $K^+ K^-$ mass squared range  
is divided into 39 slices using 40 boundary points. 
Because the final state contains two indistinguishable
$ K^+ $ mesons, the $ K^+ K^- $ amplitudes interfere
with themselves, and this sort of interference 
allows the \swave phase to be determined. 
To test the GooFit extension to support MIPWA fits,
samples of 100K signal events are generated from a Dalitz plot PDF (``Toy MC''). 
The mass projections from the fit to a test sample can be 
seen in Fig.~\ref{fig:fitproj}. 
First, the fit quality of the MIPWA method is tested
by comparing the fitted \{$a_k, \phi_k$\} % {$a^k_0, \phi^k_0$} 
values with the inputs. 
In each fit iteration, 
the Dalitz plot normalisation method (``{\tt DalitzPlotPdf::normalise()}'')
is called so that the integral of the total PDF is equal to one. 
The normalisation integral is calculated numerically based on
evenly distributed grid points in the Dalitz plot plane. 
Figure~\ref{fig:swaveshift} shows the improvement in the fit quality
as the normalisation grid spacing is reduced. 
Although the finer granularity 
increases numbers of calculations, the high
speed of the GPUs makes this problem tractable.

\begin{center}
\begin{table}[h]
\caption{\label{tbl:gpuperformance}GPU performance of the MIPWA fit.}
\footnotesize\rm
\centering
\begin{tabular}{c|c|c|c|c}\hline
    Platform & GPU Model & Chip & CUDA cores & Run time (sec.)\\\hline
    tWorkstation & Tesla K40c & GK110BGL & 2880 & 76 \\
    Desktop PC & GeForce GTX 980 & 2nd gen. Maxwell (GM204) &  2048 & 67\\ 
    Laptop ASUS N56V & GeForce GT 650M & GK107 & 384 & 179\\ \hline
\end{tabular}
\end{table}
\end{center}

As shown in Table~\ref{tbl:gpuperformance}, the GooFit
     performance on 
   three different GPU platforms has been measured 
   by running the MIPWA fit over the same toy MC sample of 100 K events (see Fig.~\ref{fig:fitproj} for the fit projections). 
 For comparison, the MIPWA fit using an older code 
 with the same functionality that runs on one CPU core  
 takes about eight hours to complete. 
 Perhaps surprisingly, an older generation mobile GPU 
(a GeForce GT 650M with 384 cores) provides excellent performance; 
a newer HPC GPU board (a Tesla K40c with 2880 cores) provides better 
performance, but not in proportion to the number of cores; 
a high-end gamer board (a GeForce GTX 980) provides the best 
performance, albeit by a small margin.
Based on other studies with GooFit, significantly 
better performance on the new P100 boards is anticipated which utilise 
NVIDIA's new Pascal GPU architecture. 

\begin{center}
\begin{table}[h]
\caption{\label{tbl:cpuplatforms}Specifications of the testing CPU platforms. 
    Asterisks next to the number of cores
indicate hyperthreading - two virtual processors per physical core.}
\footnotesize\rm
\centering
\begin{tabular}{c|cccc}\hline
    Name & Chip type  & \# of Cores & Clock [GHz] & RAM [GiB]\\\hline
    Cerberus & Intel Xeon E5520 & 8* & 2.27 & 24 \\
    Goofy & Intel Xeon CPU E5-2680 v3 & 24* & 2.50 & 120\\ \hline
\end{tabular}
\end{table}
\end{center}

\begin{figure}%[h]
\centering
%\begin{minipage}{14pc}
\includegraphics[width=0.95\textwidth]{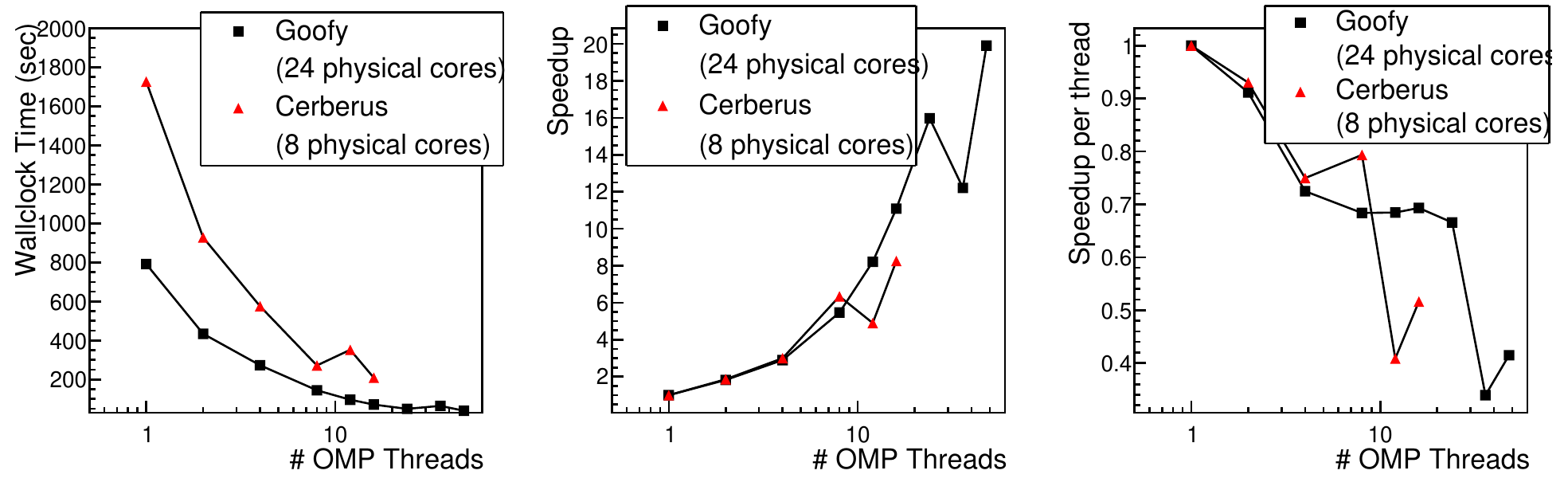}
    \caption{\label{fig:ompfit}Timing results and speedups for the test MIPWA 
    fit. Tested on two different CPU 
    platforms as shown in Table~\ref{tbl:cpuplatforms}: 
    Goofy (Intel Xeon CPU E5-2680 v3 x 2)
    and Cerberus (Intel Xeon CPU E5520 x 2).}
\end{figure}

In addition to timing GooFit's new MIPWA performance on 
GPUs using Thrust's CUDA
backend,  its performance on  
two different CPU platforms is timed using Thrust's OpenMP backend, 
as shown in Fig.~\ref{fig:ompfit}. 
With two Intel Xeon E5-2680 v3 CPUs, the fit uses 791 seconds for one OpenMP 
thread, 50 seconds for 24 threads. 
The speedup is almost linear with the increase of the number of OpenMP threads,
up until the number of threads equals the numbers of physical cores.  

\section{Summary}
This paper describes an extension of GooFit to support MIPWA fits for three-body decays, 
and have achieved speedups of
\textcolor{black}{a few hundred} by using GPUs. 
The main branch of GooFit's source code is in a GooFit repository at \url{https://github.com/GooFit/GooFit}, while 
    the updated code with MIPWA support is in a personal GooFit branch at \url{https://github.com/liang-sun/GooFit}.

\subsection{Acknowledgments}
This work was performed with support from NSF Award PHY-1414736.
NVidia provided K40 GPUs for our use through its University 
Partnership program.
The Ohio Supercomputer Center made their
``Oakley'' computer farm available for development, for testing, and for GooFit outreach workshops.

\end{document}